\documentclass[graybox]{svmult}

\usepackage{mathptmx}       
\usepackage{helvet}         
\usepackage{courier}        
\usepackage{type1cm}        
\usepackage{makeidx}         
\usepackage{graphicx}        
\usepackage{multicol}        
\usepackage[bottom]{footmisc}

\makeindex             

\usepackage{latexsym,amsmath,amssymb,amscd,eucal}
\usepackage{xypic}
\usepackage{enumerate}

\def\harr#1#2{\smash{\mathop{\hbox to .3in{\rightarrowfill}}
 \limits^{\scriptstyle#1}_{\scriptstyle#2}}}

\def\appendix#1{\addtocounter{section}{1}\setcounter{equation}{0}
\renewcommand{\thesection}{\Alph{section}}
\section*{Appendix \thesection\protect\indent \parbox[t]{11.715cm} {#1}}
\addcontentsline{toc}{section}{Appendix \thesection\ \ \ #1} }
\newcommand{\eq}{\begin{equation}}
\newcommand{\eqend}{\end{equation}}

\newbox\ncintdbox \newbox\ncinttbox
\setbox0=\hbox{$-$} \setbox2=\hbox{$\displaystyle\int$}
\setbox\ncintdbox=\hbox{\rlap{\hbox to
\wd2{\hskip-.125em\box2\relax \hfil}}\box0\kern.1em}
\setbox0=\hbox{$\vcenter{\hrule width 4pt}$}
\setbox2=\hbox{$\textstyle\int$}
\setbox\ncinttbox=\hbox{\rlap{\hbox to
\wd2{\hskip-.175em\box2\relax \hfil}}\box0\kern.1em}

\def\Z{{\mathbb{Z}}}

\def\be{\begin{equation}}
\def\ee{\end{equation}}
\def\bea{\begin{eqnarray}}
\def\eea{\end{eqnarray}}
\def\bd{\begin{displaymath}}
\def\ed{\end{displaymath}}

\DeclareFontFamily{U}{rsf}{}
\DeclareFontShape{U}{rsf}{m}{n}{
  <5> <6> rsfs5 <7> <8> <9> rsfs7 <10-> rsfs10}{}
\DeclareMathAlphabet\Scr{U}{rsf}{m}{n}

\def\cR{{\Scr R}}
\def\cS{{\Scr S}}

\def\cD{{\Scr D}}

\def\cH{{\Scr H}}

\def\cF{{\Scr F}}

\makeatletter
\newdimen\normalarrayskip              
\newdimen\minarrayskip                 
\normalarrayskip\baselineskip
\minarrayskip\jot
\newif\ifold             \oldtrue            
\def\arraymode{\ifold\relax\else\displaystyle\fi} 
\def\@arrayskip{\ifold\baselineskip\z@\lineskip\z@
     \else
     \baselineskip\minarrayskip\lineskip2\minarrayskip\fi}
\def\@arrayclassz{\ifcase \@lastchclass \@acolampacol \or
\@ampacol \or \or \or \@addamp \or
   \@acolampacol \or \@firstampfalse \@acol \fi
\edef\@preamble{\@preamble
  \ifcase \@chnum
     \hfil$\relax\arraymode\@sharp$\hfil
     \or $\relax\arraymode\@sharp$\hfil
     \or \hfil$\relax\arraymode\@sharp$\fi}}
\def\@array[#1]#2{\setbox\@arstrutbox=\hbox{\vrule
     height\arraystretch \ht\strutbox
     depth\arraystretch \dp\strutbox
     width\z@}\@mkpream{#2}\edef\@preamble{\halign \noexpand\@halignto
\bgroup \tabskip\z@ \@arstrut \@preamble \tabskip\z@ \cr}%
\let\@startpbox\@@startpbox \let\@endpbox\@@endpbox
  \if #1t\vtop \else \if#1b\vbox \else \vcenter \fi\fi
  \bgroup \let\par\relax
  \let\@sharp##\let\protect\relax
  \@arrayskip\@preamble}
\makeatother

\newcommand{\beq}{\begin{eqnarray}}
\newcommand{\eeq}{\end{eqnarray}}

\newcommand{\G}{\Gamma}

\def\appendix#1{\addtocounter{section}{1}\setcounter{equation}{0}
\renewcommand{\thesection}{\Alph{section}}
\section*{Appendix \thesection. #1}
\addcontentsline{toc}{section}{Appendix \thesection\ \ \ #1} }

\numberwithin{equation}{section}

\begin{document}

\title*{On Partition Functions of Hyperbolic Three-Geometry and
Associated Hilbert Schemes}
\author{A.~A. Bytsenko and E. Elizalde}
\institute{A.~A. Bytsenko \at Departamento de F\'{\i}sica, Universidade Estadual de
Londrina, Caixa Postal 6001, Londrina-Paran\'a, Brazil, \email{abyts@uel.br}
\and E. Elizalde \at Consejo Superior de Investigaciones Cient\'{\i}ficas (ICE/CSIC)
and Institut d'Estudis Espacials de Catalunya (IEEC),
Campus UAB, Facultat de Ci\`encies, Torre C5-Par-2a, 08193 Bellaterra (Barcelona) Spain \email{elizalde@ieec.uab.es; elizalde@math.mit.edu}}
%
%
\maketitle

\abstract*{Highest-weight representations of infinite dimensional Lie algebras and Hilbert schemes
of points are considered, together with the applications of these concepts to partition
functions, which are most useful in physics.
Partition functions (elliptic genera) are conveniently transformed into product expressions,
which may inherit the homology properties of appropriate (poly)graded Lie algebras.
Specifically, the role of (Selberg-type) Ruelle spectral functions of hyperbolic geometry in the calculation of partition functions and associated $q$-series are discussed.
Examples of these connection in quantum field theory are considered (in particular, within the AdS/CFT correspondence), as the AdS$_{3}$ case where one has Ruelle/Selberg spectral functions, whereas on the CFT side, partition functions and modular forms arise. These objects are here shown to have a common background, expressible in terms of Euler-Poincar\'e and Macdonald identities, which, in turn, describe homological aspects of (finite or infinite) Lie algebra representations. Finally, some other applications of modular forms and spectral functions (mainly related with the congruence subgroup of $SL(2, {\mathbb Z})$) to partition functions, Hilbert schemes of points, and symmetric products
are investigated by means of homological and K-theory methods.}

\abstract{Highest-weight representations of infinite dimensional Lie algebras and Hilbert schemes
of points are considered, together with the applications of these concepts to partition
functions, which are most useful in physics.
Partition functions (elliptic genera) are conveniently transformed into product expressions,
which may inherit the homology properties of appropriate (poly)graded Lie algebras.
Specifically, the role of (Selberg-type) Ruelle spectral functions of hyperbolic geometry in the calculation of partition functions and associated $q$-series are discussed.
Examples of these connection in quantum field theory are considered (in particular, within the AdS/CFT correspondence), as the AdS$_{3}$ case where one has Ruelle/Selberg spectral functions, whereas on the CFT side, partition functions and modular forms arise. These objects are here shown to have a common background, expressible in terms of Euler-Poincar\'e and Macdonald identities, which, in turn, describe homological aspects of (finite or infinite) Lie algebra representations. Finally, some other applications of modular forms and spectral functions (mainly related with the congruence subgroup of $SL(2, {\mathbb Z})$) to partition functions, Hilbert schemes of points, and symmetric products
are investigated by means of homological and K-theory methods.}


\section{Introduction}
This paper deals with highest-weight representations of infinite dimensional Lie algebras and Hilbert schemes of points, and considers applications of these concepts to partition functions, which are ubiquitous and very useful in physics. The role of (Selberg-type) Ruelle spectral functions of hyperbolic geometry in the calculation of the partition functions and associated $q$-series will be discussed. Among mathematicians, partition functions are commonly associated with new mathematical invariants for spaces, while for physicists they are one-loop partition functions of quantum field theories. In fact, partition functions (elliptic genera) can be conveniently transformed into product expressions, which may inherit the homological properties of appropriate (poly)graded Lie algebras. In quantum field theory the connection referred to above is particularly striking in the case of the so-called AdS/CFT correspondence. In the anti de Sitter space AdS$_{3}$, for instance, one has Ruelle/Selberg spectral functions, whereas on the conformal field theory (CFT) side, on the other hand, one encounters partition functions and modular forms. What we are ready to show here is that these objects do have a common background, expressible in terms of Euler-Poincar\'e and Macdonald identities which, in turn, describe homological aspects of (finite or infinite) Lie algebra representations, what is both quite remarkable and useful.

Being more specific, we will be dealing in what follows with applications of modular forms and spectral functions (mainly related to the congruence subgroup of $SL(2, {\mathbb Z})$) to partition functions, Hilbert schemes of points, and symmetric products. Here are the contents of the paper. In Sect.~\ref{Classification}  we shortly discuss the case of two-geometries and then present Thurston's list of three-geometries. This list has been organized in terms of the corresponding compact stabilizers being isomorphic to ${SO}(3)$, ${SO}(2)$, or the trivial group $\{1\}$, respectively. The analogue list of four-geometries and the corresponding stabilizer subgroups are considered in Sect.~2.3. Special attention is paid to the important case of hyperbolic three-geometry.

In Sect.~\ref{Ray-Singer} we introduce the Petterson-Selberg and the Ruelle spectral functions of hyperbolic three-geometry. Later, in Sect.~\ref{Hyperbolic}, we consider examples for which we explicitly show that the respective partition functions can be written in terms of Ruelle's spectral functions associated with $q$-series, although the hyperbolic side remains still to be explored.
In Sect.~\ref{Symmetric} we briefly explain the relationship existing between the Heisenberg algebra and its representation, and with the Hilbert scheme of points in Sect.~\ref{HilbertSchemes}. This allows to construct a representation of products of Heisenberg and Clifford algebras on the direct sum of homology groups of all components associated with schemes. Hilbert schemes of points of surfaces are discussed in Sect.~\ref{Variety}; we rewrite there the character formulas and G\"{o}ttsche's formula in terms of Ruelle's spectral functions.

In Sect.~\ref{K-theory} we pay attention to the special case of algebraic structures of the K-groups $K_{\widetilde{H}\Gamma_N}(X^N)$ of $\Gamma_N$-equivariant Clifford supermodulus on $X^N$, following the lines of \cite{Wang}. This case is important since the direct sum $\cF_{\Gamma}^{-}(X)= \oplus_N^{\infty}K_{\widetilde{H}\Gamma_N}(X^N)$ naturally carries a Hopf algebra structure, and it is isomorphic to the Fock space of a twisted Heisenberg superalgebra with $K_{\widetilde{H}\Gamma_N}(X)\cong K_{\Gamma}(X)$.
In terms of the Ruelle spectral function we represent the dimension of a direct sum of the equivariant K-groups (related to a suitable supersymmetric algebra). We analyze elliptic genera for generalized wreath and symmetric products on $N$-folds; these cases are examples of rather straightforward applications of the machinery of modular forms and spectral functions discussed above.
Finally, in the conclusions, Sect.~\ref{Conclusions}, we briefly outline some issues and further perspectives for the analysis of partition functions in connection with deformation quantization.

\section{Classification of low-dimensional geometries}
\label{Classification}
The problem of the classification of geometries is  most important
in complex analysis and in mathematics as a whole, and also plays a
fundamental role in physical theories.
Indeed, in quantum field theory functional integration
over spaces of metrics can be separated into an integration over
all metrics for some volume of a definite topology, followed
by a sum over all topologies.
But even for low-dimensional spaces (say, for example, the
three-dimensional case) of fixed topology, the moduli space of all
metric diffeomorphisms is infinite dimensional. And this leads
back to the deep mathematical task associated with the classification
problem. In this section we present a brief discussion of
the classification (uniformization) issue and of the sum over the topology for
low-dimensional cases, allong the lines of \cite{Filipkiewicz,Wall,Bytsenko2}.

All curves of genus zero can be uniformized
by rational functions, those of genus one  by
elliptic functions, and those of genus higher than one
by meromorphic functions, defined on proper open subsets
of ${\mathbb C}$. These results, due to
Klein, Poincar\'e and Koebe, are among the deepest achievements in
mathematics. A complete solution of the uniformization
problem has not yet been obtained (with the exception of the
one-dimensional complex case). However, there have been essential
advances in this problem, which have brought to the foundation of
topological methods, covering spaces, existence theorems for partial
differential equations, existence and distortion theorems for
conformal mappings, etc.

\noindent
{\bf Three-geometries.}
According to Thurston's conjecture \cite{Thurston}, there are eight model spaces in three dimensions:
\begin{equation*}
{X}= G/\Gamma =\left\{\!\begin{array}{ll}
{\mathbb R}^3\, {\rm (Euclidean \,\,\,\, space)}\,,\,\,\,\,
{S}^3\, {\rm (spherical \,\,\, space)}\,, \,\,\,\,
{H}^3 \,{\rm (hyperbolic \,\,\, space)}
\\
{H}^2\times{\mathbb R}\,,\,\,\,\,
{S}^2\times {\mathbb R}\,,\,\,\,\,
\widetilde{{S}{L}(2,{\mathbb R})}\,,\,\,\,\,
{N}il^3\,,\,\,\,\,
{S}ol^3\,
\end{array}\!\right\}
\end{equation*}
An important remark is in order.
This conjecture follows from considering the identity component of the isotropy group, $\Gamma \equiv \Gamma_x$ of $X$, through a point, $x$. $\Gamma$ is a compact, connected Lie group, and
one must distinguish the three different cases: $\Gamma = {SO}(3),\, {SO}(2)$
and $\{ 1 \}$.
\begin{enumerate}
\item{}
$\Gamma = {SO}(3)$. In this case the space $X$
has constant curvature: ${\mathbb R}^3,\, {S}^3$ {\rm (}modeled on ${\mathbb R}^3${\rm )} or ${H}^3$ {\rm (}which can be modeled on the half-space ${\mathbb R}^2\times {\mathbb R}^{+}${\rm )}.
\item{}
$\Gamma = {SO}(2)$. In this case there is a one-dimensional subspace of $TX$ that is invariant under $\Gamma$, which has a complementary plane field ${P}_x$. If the plane field ${P}_x$ is integrable, then $X$ is a product ${\mathbb R}\times {S}^2$ or ${\mathbb R}\times {H}^2$. If the plane field ${P}_x$ is non-integrable, then $X$ is a non-trivial fiber bundle with fiber $S^1$:
$S^1 \hookrightarrow X \twoheadrightarrow \Sigma_{g\geq 2}$
{\rm (}$\widetilde{{S}{L}(2,{\mathbb R})}-$geometry{\rm )},\,
$\Sigma_g$ stands for a surface of genus $g$,\,
$S^1 \hookrightarrow X \twoheadrightarrow {T}^2$
{\rm (}${N}il^3$-geometry{\rm )}\, or
$S^1 \hookrightarrow X \twoheadrightarrow {S}^2$\,
{\rm (}${S}^3-$geometry{\rm )}.
\item{}
$\Gamma = \{1\}$. In this case we have the three-dimensional Lie groups:
$\widetilde{{S}{L}(2,{\mathbb R})}\,,\,
{N}il^3$, and ${S}ol^3$\,.
\end{enumerate}

The first five geometries are familiar objects, so we briefly discuss the last three of them. The group $\widetilde{{S}{L}(2,{\mathbb R})}$ is the universal covering of ${S}{L}(2,{\mathbb R})$, the three-dimensional Lie group of all $2\times 2$ real matrices with determinant equal to 1. The geometry of ${N}il$ is the three-dimensional Lie group of all $3 \times 3$ real upper triangular matrices endowed with
ordinary matrix multiplication. It is also known as the nilpotent Heisenberg group. The geometry of ${S}ol$ is the three-dimensional (solvable) group.
\\

\noindent
{\bf Four-geometries.}
The list of Thurston three-geometries has been organized in terms of the compact stabilizers $\Gamma$. The analogue list of
four-geometries can also be organized using connected groups of isometries only (Table 2).
\begin{table}\label{4-spaces}
\begin{center}
\begin{tabular}
{l l}
Table 2. List of the four-geometries
\\
\\
\hline
\\
Stabilizer-subgroup $\Gamma$ & Space \, $X$ \\
\\
\hline
\\
${SO}(4)$ & ${\mathbb R}^4$,\,\, ${S}^4,\,\, {H}^4$ \\
${U}(2)$ & ${\mathbb C} P^2,\,{\mathbb C}{H}^2$ \\
${SO}(2)\times {SO}(2)$ & ${S}^2\times {\mathbb R}^2,\,\,
{S}^2\times
{S}^2,\,\,{S}^2\times {H}^2,\,\, {H}^2\times
{\mathbb R}^2,\,\,{H}^2\times {H}^2$ \\
${SO}(3)$ & ${S}^3\times {\mathbb R},\,\, {H}^3\times
{\mathbb R}$ \\
${SO}(2)$ & ${N}il^3\times
{\mathbb R},\,\,{\widetilde{{PSL}}}(2,{\mathbb R})\times {\mathbb R},\,\,{S}ol^4$ \\
${S}^1$ & $F^4$ \\
{\rm trivial} & ${N}il^4,\,\, {S}ol^4_{m,n}$ (including
${S}ol^3\times {\mathbb R}), {S}ol^4_1$ \\
\\
\hline
\end{tabular}
\end{center}
\end{table}
Here we have the four irreducible four-dimensional Riemannian symmetric spaces: sphere ${S}^4$,
hyperbolic space ${H}^4$, complex projective space ${\mathbb C}
P^2$, and complex hyperbolic space ${\mathbb C} {H}^2$ (which we
may identify with the open unit ball in ${\mathbb C}^2$, with
an appropriate metric). The other cases are more specific and
 we shall illustrate them for the sake of completeness only.

The nilpotent Lie group ${N}il^4$ can be presented as the split
extension ${\mathbb R}^3 \rtimes_U {\mathbb R}$ of
${\mathbb R}^3$ by ${\mathbb R}$, where the real 3-dimensional
representation $U$ of $\mathbb R$ has the form $U(t)=\exp (tB)$ with
$
B=\left( \begin{array}{ccc} 0 & 1 & 0\\ 0 & 0 & 1\\ 0
& 0 & 0 \end{array} \right)\,.
$
In the same way, the soluble Lie groups
${S}ol_{m,n}^4= {\mathbb R}^3 \rtimes _{T_{m,n}}{\mathbb R}$,
on real three-dimensional representations $T_{m,n}$ of ${\mathbb R}$, $T_{m,n}(t)=\exp (tC_{m,n})$, where
$
C_{m,n}=
{\rm diag}(\alpha, \beta, \gamma)
$
and $ \alpha+\beta+\gamma=0 $ for $\alpha>\beta>\gamma$. Furthermore
$e^{\alpha}$, $e^{\beta}$ and $e^{\gamma}$ are the roots of
$\lambda^3-m\lambda^2+n\lambda-1=0$, with $m,\;n$ positive integers.
If $m=n$, then $\beta=0$ and  ${S}ol_{m,n}^4= {S}ol^3\times {\mathbb R}
$. In general, if $C_{m,n}\propto C_{m',n'}$, then ${S}ol_{m,n}^4\cong
{S}ol_{m',n'}^4$. It gives infinitely many classes of equivalence.
When $m^2n^2+18=4(m^3+n^3)+27$, one has a new
geometry, ${S}ol^4_0$, associated with the group ${SO}(2) $ of isometries rotating the first two coordinates. The soluble group ${S}ol^4_1$, is most conveniently represented as the matrix group
$
\left\{\left(\begin{array}{ccc} 1 & b & c\\ 0 &
\alpha & a\\ 0 & 0 & 1 \end{array} \right)\,
: \,\,\,\alpha, a, b, c \in {\mathbb R}, \alpha> 0 \right\}\,.
$
Finally, the geometry $F^4$ is associated with the isometry group
${\mathbb R}^2 \rtimes {PSL}(2,{\mathbb R})$ and stabilizer ${SO}(2)$. Here the semidirect product is taken with respect to the action of the group
${PSL}(2,{\mathbb R})$ on ${\mathbb R}^2$.
The space $F^4$ is diffeomorphic to ${\mathbb R}^4$
and has alternating signs in the metric. A connection of these geometries with complex and K\"ahlerian structures (preserved by the stabilizer $\Gamma$) can be found in \cite{Wall}.

We conclude this section with some comments.
In two-dimensional quantum theory it is customary to perform the sum over all topologies. Then, any functional integral of fixed genus $g$ can be written in the form:
$
\int [Dg] =\sum_{g=0}^ {\infty}\int_{(\rm fixed\,\,genus)}[Dg]
\,.
$
A necessary first step to implement this in the three-dimensional
case is the classification of all possible three-topologies
{\rm (}by Kleinian groups{\rm )}. Provided Thurston's conjecture is true, every compact closed
three-dimensional manifold can be represented as
$
\bigcup_{\ell=1}^{\infty}G_{n_{\ell}}/\Gamma_{n_{\ell}}\,,
$
where $n_{\ell} \in (1,...,8)$ represents one of the eight geometries,
and $\Gamma$ is the {\rm (}discrete{\rm )} isometry group of the corresponding geometry.  It has to be noted that gluing the above
geometries, characterizing different coupling  constants by a
complicated set of moduli, is a very difficult task.
Perhaps this could be done, however, with a bit of luck, but the more important contribution to the vacuum persistence amplitude should be given  by the compact hyperbolic geometry, the other geometries appearing only for a small number of exceptions {\rm \cite{Besse}}.
Indeed, many three-manifolds are hyperbolic {\rm (}according to a famous theorem by Thurston {\rm \cite{Thurston}}{\rm )}. For example, the complement of
a knot in ${S}^3$ admits a hyperbolic structure unless it is a
torus or satellite knot. Moreover, according to the Mostow Rigidity Theorem
{\rm \cite{Mostow}}, any geometric invariant of a hyperbolic
three-manifold is a topological invariant. Our special interest here is directed towards hyperbolic spaces. Some examples of partition functions and elliptic genera written in term of spectral functions of ${H}^3$ spaces and their quotients by a subgroup of the isometry group
${PSL}(2, {\mathbb C})\equiv {SL}(2, {\mathbb C})/\{\pm 1\}$ can be found in {\rm \cite{BonoraBytsenko,Bonora}}.

\section{The Ray-Singer norm. Hyperbolic three-geometry}
\label{Ray-Singer}
If ${\mathfrak L}_p$ is a self-adjoint Laplacian on $p$-forms then
the following results hold. There exists $\varepsilon,\delta >0$
such that for $0<t<\delta$ the heat kernel expansion for
Laplace operators on a compact manifold $X$ is given by
$
{\rm Tr}\left(e^{-t{\mathfrak L}_p}\right)=
\sum_{0\leq \ell\leq \ell_0} a_\ell
({\mathfrak L}_p)t^{-\ell}+ {O}(t^\varepsilon).
$
The zeta function of ${\mathfrak L}_p$ is the Mellin transform
\begin{equation}
\zeta(s|{\mathfrak L}_p)=
{\mathfrak M}\left[{\rm Tr}e^{-t{\mathfrak L}_p}\right]
=[\Gamma(s)]^{-1}\int_{{\mathbb R}_{+}}
{\rm Tr}e^{-t{\mathfrak L}_p}t^{s-1}dt
\mbox{.}
\end{equation}
This function equals ${\rm Tr}\left({\mathfrak L}_p^{-s}\right)$ for
$s> (1/2)\, {\rm dim}\,X$.
Let $\chi$ be an orthogonal representation of $\pi_1(X)$.
Using the Hodge decomposition, the vector space $H(X;\chi)$
of twisted cohomology classes can be embedded into
$\Omega(X;\chi)$ as the space of harmonic forms. This embedding
induces a norm $|\cdot|^{RS}$ on the determinant line ${\rm det}H(X;\chi)$. The
Ray-Singer norm $||\cdot||^{RS}$ on ${\rm det}H(X;\chi)$ is defined
by \cite{Ray}
\begin{equation}
||\cdot||^{RS}\stackrel{def}=|\cdot|^{RS}\prod_{p=0}^{{\rm dim}X}
\left[\exp\left(-\frac{d}{ds}
\zeta (s|{\mathfrak L}_p)|_{s=0}\right)\right]^{(-1)^pp/2}
\mbox{,}
\end{equation}
where the zeta function $\zeta (s|{\mathfrak L}_p)$ of the Laplacian
acting on the space of
$p$-forms orthogonal to the harmonic forms has been used. For a
closed connected orientable smooth manifold of odd dimension,
and for the Euler structure
$\eta\in {\rm Eul}(X)$, the Ray-Singer norm of its cohomological
torsion $\tau_{an}(X;\eta)=\tau_{an}(X)\in {\rm det}H(X;\chi)$ is
equal to the positive square root of the absolute value of the monodromy of $\chi$
along the characteristic class $c(\eta)\in H^1(X)$ \cite{Farber}:
$||\tau_{an}(X)||^{RS}=|{\rm det}_{\chi}c(\eta)|^{1/2}$.
And in the special case where the flat bundle $\chi$ is acyclic (namely the vector space $H^q(X; \chi)$ of twisted cohomology is zero), we have
\begin{equation}
\left[\tau_{an}(X)\right]^2
=|{\rm det}_{\chi}c(\eta)|
\prod_{p=0}^{{\rm dim}X}\left[\exp\left(-\frac{d}{ds}
\zeta (s|{\mathfrak L}_p)|_{s=0}\right)\right]^{(-1)^{p+1}p}
\mbox{.}
\label{RS}
\end{equation}

\subsection{Spectral functions of hyperbolic three-geometry}
\label{Hyperbolic}
For a  closed oriented hyperbolic three-manifold of the form
$X_{\Gamma} = {H}^3/\Gamma$ and for acyclic
$\chi$ the analytic torsion reads \cite{Fried,Bytsenko3,Bytsenko4}:
$[\tau_{an}(X_{\Gamma})]^2={\cR}(0)$, where
${\cR}(s)$ is the Ruelle function.\footnote{Vanishing theorems for the type $(0,q)$ cohomology of locally symmetric spaces can be found in \cite{Williams}. Again, if $\chi$ is acyclic ($H(X;\chi)=0$) the Ray-Singer norm (\ref{RS}) is a topological invariant: it does not depend on the choice of the metric on $X$ and $\chi$, used in the construction. If $X$ is a complex manifold (smooth $C^{\infty}$-manifold, or topological space) then
${\mathbb E}\rightarrow X$ is the induced complex (or smooth, or continuous) vector bundles. We write $H^{p,q}(X;{\mathbb E})\simeq H^{0,q}(X;\Lambda^{p,0}X\otimes {\mathbb E})$ holonomic vector bundles $\Lambda^{p,0}X\rightarrow X$ (see \cite{Williams} for details).
}
A Ruelle type zeta function, for ${\rm Re}\,s$
large, can be defined as the product over prime closed geodesics $\gamma$
of factors ${\rm det}(I-\xi(\gamma)e^{-s\ell(\gamma)})$,
where $\ell(\gamma)$ is the length of $\gamma$, and can be
continued meromorphically to the entire complex plane
$\mathbb C$ \cite{Deitmar}.
The function ${\cR}(s)$ is an alternating
product of more complicated factors, each of which is a Selberg zeta function
$Z_{\Gamma}(s)$. The relation between the Ruelle and Selberg functions is:
\begin{equation}
{\cR}(s)=\prod_{p=0}^{{\rm dim}X-1}Z_{\Gamma}
(p+s)^{(-1)^j}
\mbox{.}
\end{equation}
The Ruelle function associated with closed oriented
hyperbolic three-manifold $X_{\Gamma}$ has the
form:
${\cR}(s)=Z_{\Gamma}(s)Z_{\Gamma}(2+s)
/Z_{\Gamma}(1+s)$.

We would like here to shed light on some aspects of the so-called
AdS$_3$/CFT$_2$ correspondence, which plays a very important role in quantum field theory. Indeed, it is known that the geometric structure of three-dimensional gravity allows for exact computations, since its Euclidean counterpart is locally isomorphic to constant curvature hyperbolic space. Because of the AdS$_3$/CFT$_2$ correspondence, one expects a correspondence between spectral functions related to Euclidean AdS$_3$ and modular-like functions {\rm (}Poincar\'e series{\rm )}.\footnote{The modular forms in question are the forms for the congruence subgroup of $SL(2, {\mathbb Z})$, which is viewed as the group that leaves fixed one of the three non-trivial spin structures on an elliptic curve.}
One assumes this correspondence to occur when the arguments of the spectral functions take values on a Riemann surface, viewed as the conformal boundary of AdS$_3$. According to the holographic principle,  strong ties exist between certain field theory quantities on
the bulk of an AdS$_3$ manifold and related quantities on its boundary at infinity. To be more precise, the classes of Euclidean AdS$_3$ spaces are quotients of the real hyperbolic space by a discrete group {\rm (}a Schottky group{\rm )}. The boundary of these spaces can be compact oriented surfaces with conformal structure {\rm (}compact complex algebraic curves{\rm )}. A general formulation of the ``Holography Principle'' states that there is a correspondence between a certain class of fields, their properties and their correlators in the bulk space, where gravity propagates, and a class of primary fields, with their properties and correlators of the conformal theory on the boundary. More precisely, the set of scattering poles in $3D$ coincides with the zeroes of a Selberg-type spectral function {\rm \cite{Perry,Bytsenko4}}. Thus, encoded on a Selberg function is the spectrum of a three-dimensional model.
In the framework of this general principle, we would like to illustrate the correspondence between spectral functions of hyperbolic three-geometry {\rm (}its spectrum being encoded in the Petterson-Selberg spectral functions{\rm )} and Poincar\'e series associated with the conformal structure in two dimensions.

Let us consider a three-geometry with an orbifold description $H^3/\Gamma$. The complex unimodular group $G=SL(2, {\mathbb C})$
acts on the real hyperbolic three-space $H^3$ in a standard way, namely for $(x,y,z)\in H^3$ and $g\in G$, one has
$g\cdot(x,y,z)= (u,v,w)\in H^3$. Thus, for $r=x+iy$,\,
$g= \left[ \begin{array}{cc} a & b \\ c & d \end{array} \right]$,
$
u+iv = [(ar+b)\overline{(cr+d)}+ a\overline{c}z^2]\cdot
[|cr+d|^2 + |c|^2z^2]^{-1},\,
w = z\cdot[
{|cr+d|^2 + |c|^2z^2}]^{-1}\,.
$
Here the bar denotes complex conjugation. Let $\Gamma \in G$ be the discrete group of $G$ be
defined as
\begin{eqnarray}
\Gamma & = & \{{\rm diag}(e^{2n\pi ({\rm Im}\,\tau + i{\rm Re}\,\tau)},\,\,  e^{-2n\pi ({\rm Im}\,\tau + i{\rm Re}\,\tau)}):
n\in {\mathbb Z}\}
= \{{\mathfrak g}^n:\, n\in {\mathbb Z}\}\,,
\nonumber \\
{\mathfrak g} & = &
{\rm diag}(e^{2\pi ({\rm Im}\,\tau + i{\rm Re}\,\tau)},\,\,  e^{-2\pi ({\rm Im}\,\tau + i{\rm Re}\,\tau)})\,.
\end{eqnarray}
One can define a Selberg-type zeta function for the group
$\Gamma = \{{\mathfrak g}^n : n \in {\mathbb Z}\}$ generated by a single hyperbolic element of the form ${\mathfrak g} = {\rm diag}(e^z, e^{-z})$, where $z=\alpha+i\beta$ for $\alpha,\beta >0$. In fact, we will take
$\alpha = 2\pi {\rm Im}\,\tau$, $\beta= 2\pi {\rm Re}\,\tau$. For the standard action of $SL(2, {\mathbb C})$ on $H^3$, one has
\begin{equation}
{\mathfrak g}
\left[ \begin{array}{c} x \\ y\\ z \end{array} \right]
=
\left[\begin{array}{ccc} e^{\alpha} & 0 & 0\\ 0 & e^{\alpha} & 0\\ 0
& 0 & \,\,e^{\alpha} \end{array} \right]
\left[\begin{array}{ccc} \cos(\beta) & -\sin (\beta) & 0\\
\sin (\beta) & \,\,\,\,\cos (\beta) & 0
\\ 0 & 0 & 1 \end{array} \right]
\left[\begin{array}{c} x \\ y\\ z \end{array} \right]
\,.
\end{equation}
Therefore, ${\mathfrak g}$ is the composition of a rotation in ${\mathbb R}^2$, with complex eigenvalues $\exp (\pm i\beta)$, and a dilatation $\exp (\alpha)$.
The Patterson-Selberg spectral function $Z_\Gamma (s)$ is
meromorphic on $\mathbb C$ and can be attached to ${H}^3/\Gamma$. It
is given, for ${\rm Re}\, s> 0$, by the formulas \cite{Perry,Bytsenko07,Bytsenko08}
\begin{eqnarray}
Z_\Gamma(s)&:= &\prod_{k_1, k_2 \geq 0} [1-(e^{i\beta})^{k_1}(e^{-i\beta})^{k_2}e^{-(k_1+k_2+s)\alpha}]\,,
\label{zeta00}
\\
{\rm log}\, Z_{\Gamma}(s)& = &
-\frac{1}{4}\sum_{n=1}^{\infty}\frac{e^{-n\alpha(s-1)}}
{n[\sinh^2\left(\frac{\alpha n}{2}\right)
+\sin^2\left(\frac{\beta n}{2}\right)]}\,.
\label{logZ}
\end{eqnarray}
The zeros of $Z_\Gamma (s)$ are the complex numbers
$
\zeta_{n,k_{1},k_{2}} = -\left(k_{1}+k_{2}\right)+i\left(k_{1}-
k_{2}\right)\beta/\alpha+ 2\pi  in/\alpha,\,
$
$n \in {\mathbb Z}$ (for details, see \cite{Bytsenko07}).
It can be also shown that the zeta function $Z_\Gamma (s)$ is an entire function of order three and finite type. It is bounded in absolute value, for ${\rm Re}\,s\geq 0$ as well as for
${\rm Re}\,s\leq 0$, and can be estimate as follows:
\begin{equation}
|Z_\Gamma(s)| \leq
\prod_{\stackrel{k_1+k_2\leq
|s|}{}}^\infty e^{|s|\ell}
\prod_{\stackrel{k_1+k_2\geq
|s|}{}}^\infty (1- e^{(|s|-k_1-k_2)\ell})
\leq C_1e^{C_2|s|^3}\,,
\label{estimate}
\end{equation}
where $C_1$ and $C_2$ are suitable constants. The first factor on the right hand side of (\ref{estimate}) yields exponential growth, while the second factor is bounded, what proves the required growth estimate.
The spectral function $Z_\Gamma (s)$ is an entire function of order three and finite type, and can written as a Hadamard product \cite{Perry}
\begin{equation}
Z_\Gamma(s) =
e^{Q(s)}
\prod_{\zeta \in {\Sigma}}
(1-\frac{s}{\zeta})
\exp \left(
\frac{s}{\zeta} + \frac{s^2}{2\zeta^2} +
\frac{s^3}{3\zeta^3}\right)\,,
\label{Hadamard}
\end{equation}
where $\zeta \equiv \zeta_{n,k_{1},k_{2}}$ and we denote the set of such numbers by $\Sigma$, $Q(s)$ being a polynomial of degree at most three. It follows from the Hadamard product representation of $Z_\Gamma (s)$ (\ref{Hadamard}) that
\begin{equation}
\frac{d}{ds}{\rm log}\,Z_\Gamma (s) =
\frac{d}{ds} Q(s) + \sum_{\zeta \in \Sigma}
\frac{(s/\zeta)^3}{s-\zeta}\,.
\end{equation}
Let us define $\Xi(y \pm i\xi):= (d/ds){\rm log}Z_\Gamma (s)$
for $s = y \pm i\xi$. Then
\begin{equation}
\Xi (y \pm i\xi)
= \frac{d}{ds}Q(s= y \pm i\xi)
+ i^{-1}\sum_{y \pm i\varepsilon \in \Sigma}
\frac{(y \pm i\xi)^3}
{(y \pm i\varepsilon)^3(\pm \xi- \varepsilon)}\,.
\label{Phi}
\end{equation}

\noindent {\bf Generating functions.}
Using the equality
$
\sinh^2\left(\alpha n/2\right)
+ \sin^2\left(\beta n/2\right)$ $ = |\sin(n\pi \tau)|^2 =
|1-q^n|^2/(4|q|^{n})
$
and Eq.~(\ref{logZ}), we get
\begin{eqnarray}
{\rm log}\prod_{m=\ell}^{\infty}(1- q^{m+\varepsilon})
& = &
\sum_{m=\ell}^{\infty}
{\rm log}(1 - q^{m+\varepsilon})
=
-\sum_{n=1}^{\infty}
\frac{q^{(\ell + \varepsilon)n}
(1- \overline{q}^{n})|q|^{-n}}
{4n|\sin(n\pi \tau)|^2}
\nonumber \\
&=&
{\rm log}\,\left[\frac{Z_\Gamma(\xi (1-it))}{Z_\Gamma(
\xi (1-it)+1+it)}\right]\,,
\label{modular1}
\\
{\rm log}\prod_{m=\ell}^{\infty}(1- \overline{q}^{m+\varepsilon})
& = &
\sum_{m=\ell}^{\infty}
{\rm log}(1 - {\overline q}^{m+\varepsilon})
=
-\sum_{n=1}^{\infty}
\frac{{\overline q}^{(\ell + \varepsilon)n}
(1- {q}^{n})|q|^{-n}}
{4n|\sin(n\pi \tau)|^2}
\nonumber \\
& = &
{\rm log}\,\left[\frac{Z_\Gamma(\xi (1+it))}{Z_\Gamma(
\xi (1+it)+1-it)}\right]\,,
\label{modular2}
\\
{\rm log}\prod_{m=\ell}^{\infty}(1+ {q}^{m+\varepsilon})
& = &
\sum_{m=\ell}^{\infty}
{\rm log}(1 + {q}^{m+\varepsilon})
=
-\sum_{n=1}^{\infty}
\frac{(-1)^n q^{(\ell + \varepsilon)n}
(1- {\overline q}^{n})|q|^{-n}}
{4n|\sin(n\pi \tau)|^2}
\nonumber \\
& = &
{\rm log}\,\left[\frac{Z_\Gamma(\xi (1-it) +
i\eta(\tau))}{Z_\Gamma(\xi (1+it)+1-it + i\eta(\tau))}\right]\,,
\label{modular3}
\\
{\rm log}\prod_{m=\ell}^{\infty}(1+ {\overline q}^{m+\varepsilon})
& = &
\sum_{m=\ell}^{\infty}
{\rm log}(1 + {\overline q}^{m+\varepsilon})
=
-\sum_{n=1}^{\infty}
\frac{(-1)^nq^{(\ell + \varepsilon)n}
(1- {q}^{n})|q|^{-n}}
{4n|\sin(n\pi \tau)|^2}
\nonumber \\
& = &
{\rm log}\,\left[\frac{Z_\Gamma(\xi(1+it) + i\eta(\tau))}{Z_\Gamma(\xi(1+it)+1-it + i\eta(\tau))}\right]\,,
\label{modular4}
\end{eqnarray}
where $\ell \in {\mathbb Z}_+,\, \varepsilon \in {\mathbb C}$, $t = {\rm Re}\,\tau/{\rm Im}\,\tau$, $\xi = \ell + \varepsilon$ and $\eta (\tau)= \pm(2\tau)^{-1}$.
Let us next introduce some well-known functions and their modular properties under the action of $SL(2, {\mathbb Z})$. The special cases associated with (\ref{modular1}), (\ref{modular2})
are (see \cite{Kac}):
\begin{eqnarray}
f_1(q) & = & q^{-\frac{1}{48}}\prod_{m > 0}
(1-q^{m+\frac{1}{2}})\,\, = \,\, \frac{\eta_D(q^{\frac{1}{2}})}{\eta_D(q)}\,,
\\
f_2(q) & = & q^{-\frac{1}{48}}\prod_{m > 0}
(1+q^{m+\frac{1}{2}})\,\, = \,\, \frac{\eta_D(q)^2}{\eta_D(q^{\frac{1}{2}})\eta_D(q^2)}\,,
\\
f_3(q) & = & \,\, \,\, q^{\frac{1}{24}}\prod_{m > 0}
(1+q^{m+1})\,\, = \,\, \frac{\eta_D(q^2)}{\eta_D(q)}\,,
\end{eqnarray}
where
$
\eta_D(q) \equiv q^{1/24}\prod_{n > 0}(1-q^{n})
$
is the Dedekind $\eta$-function. The linear span of $f_1(q), f_2(q)$
and $f_3(q)$ is $SL(2, {\mathbb Z})$-invariant \cite{Kac}
(
$
\!g\in \left[\begin{array}{cc} a & b \\ c & d \end{array}\right],\,
$
$
g \cdot f(\tau) = f\left(\frac{a\tau + b}{c\tau + d}\right)
$
).
As
$
f_1(q)\cdot f_2(q)\cdot f_3(q) = 1,
$
we get
$
\cR(s=3/2-(3/2)it)\cdot \cR(\sigma =3/2-(3/2)it+i\eta(\tau))
\cdot \cR(\sigma =2-2it+i\eta(\tau)) = 1\,.
$
For a closed oriented hyperbolic three-manifolds of the form $X= H^3/\Gamma$ (and any acyclic orthogonal representation of $\pi_1(X)$)
a set of useful generating functions is collected in Table {\rm
\ref{Table1}}.
\begin{table}\label{Table1}
\begin{center}
\begin{tabular}
{l l }
Table {}  \ref{Table1}. \,\,\,
List of generating functions
\\
\\
\hline
\\
$ \prod_{n=\ell}^{\infty}(1-q^{n+\varepsilon}) \,\,\,\, =
\left[\frac{Z_\Gamma(\xi (1-it))}
{Z_\Gamma(\xi (1 - it)+ 1+ it)}\right]\,\,\,\,\,\,\,\,\,\,\,\,
= \,\, \cR(s= \xi(1-it))$
\\
\\
$ \prod_{n=\ell}^{\infty}(1- \overline{q}^{n+\varepsilon}) \,\,\,\, =
\left[\frac{Z_\Gamma(\xi (1+it))}
{Z_\Gamma(\xi (1 + it)+ 1- it)}\right]\,\,\,\,\,\,\,\,\,\,\,\,
= \,\, \cR(\overline{s}= \xi(1+it))$
\\
\\
$
\prod_{n=\ell}^{\infty}(1+q^{n+\varepsilon}) \,\,\,\, =
\left[\frac{Z_\Gamma(\xi(1-it) + i\eta(\tau))}
{Z_\Gamma(\xi (1-it) + i\eta(\tau)+ 1+it)}\right]
= \,\, \cR(\sigma = \xi(1-it) + i\eta(\tau))
$
\\
\\
$
\prod_{n=\ell}^{\infty}(1+ \overline{q}^{n+\varepsilon}) \,\,\,\, =
\left[\frac{Z_\Gamma(\xi(1+it) + i\eta(\tau))}
{Z_\Gamma(\xi (1+it) + i\eta(\tau)+ 1-it)}\right]
= \,\, \cR(\overline{\sigma} = \xi(1+it) + i\eta(\tau))
$
\\
\\
$\prod_{n=\ell}^{\infty}(1-q^{n+ \epsilon})^n
 =
\cR(s= \xi(1-it))^\ell\prod_{n=\ell}^{\infty}
\cR(s=(n+\varepsilon +1)(1-it))
$
\\
\\
$\prod_{n=\ell}^{\infty}(1-\overline{q}^{n+\epsilon})^n
 =
\cR(\overline{s}= \xi(1+it))^\ell\prod_{n=\ell}^{\infty}
\cR(s=(n+\varepsilon +1)(1+it))
$
\\
\\
$\prod_{n=\ell}^{\infty}(1+ q^{n+\epsilon})^n
 =
\cR(\sigma = \xi(1-it)+ i\eta(\tau))^\ell
\prod_{n=\ell}^{\infty}\cR(\sigma =(n+\varepsilon +1)(1-it)+i\eta(\tau))$
\\
\\
$
\prod_{n=\ell}^{\infty}(1+\overline{q}^{n+\epsilon})^n
 =
\cR(\overline{\sigma}= \xi(1+it)+i\eta(\tau))^\ell
\prod_{n=\ell}^{\infty} \cR(\overline{\sigma}=(n+\varepsilon +1) (1+it)+i\eta(\tau))$
\\
\\
\hline
\end{tabular}
\end{center}
\end{table}

\section{Generalized symmetric products of $N$-folds}
\label{Symmetric}
\subsection{Hilbert schemes and Heisenberg algebras}
\label{HilbertSchemes}
Before entering the discussion of the main topic in this section, a short comment about Heisenberg algebras and Hilbert schemes will be in order.
Preliminary to the subject of symmetric products and their connection with spectral functions, we briefly explain the relation between the Heisenberg algebra and its representations, and the Hilbert scheme of points, mostly following the lines of \cite{Nakajima00}.

To be more specific, note that
the infinite dimensional Heisenberg algebra {\rm (}or, simply, the Heisenberg algebra{\rm )} plays a fundamental role in the representation theory of affine Lie algebras. An important representation of the Heisenberg algebra is the Fock space representation on the polynomial ring of infinitely many variables. The degrees of polynomials {\rm (}with different degree variables{\rm )} give a direct sum decomposition of the representation, which is called weight space decomposition.

The Hilbert scheme of points on a complex surface appears in algebraic geometry. The Hilbert scheme of points decomposes into infinitely many connected components according to the number of points. Betti numbers of the Hilbert scheme have been computed in {\rm \cite{Gottsche}}. The sum of the Betti numbers of the Hilbert scheme of $N$-points is equal to the dimension of the subspaces of the Fock space representation of degree $N$.
\medskip

\noindent
{\bf Algebraic preliminaries.}
Let ${R} = {\mathbb Q}[p_1, p_2, ...]$
be the polynomial ring of infinite many variables $\{p_j\}_{j=1}^{\infty}$. Define $P[j]$ as $j\partial/\partial p_j$
and $P[- j]$ as a multiplication of $p_j$ for each positive $j$. Then, the commutation relation holds:
$
[\,P[i],\, P[j]\,] = i\delta_{i+j, 0}\,{\rm Id}_{R},\,
$
$i, j \in {\mathbb Z}/\{0\}.$ We define the infinite dimensional Heisenberg algebra as the Lie algebra generated by $P[j]$ and $K$ with defining relation
\begin{equation}
[\,P[i],\, P[j]\,] = i\delta_{i+j, 0} K_{R},\,\,\,\,\,
[\,P[i],\, K\,] = 0,\,\,\,\,\, i, j \in {\mathbb Z}/\{0\}.
\end{equation}
The above ${R}$ labels the representation.
If $1\in {R}$ is the constant polynomial, then $P[i]1 = 0,\, i\in {\mathbb Z_+}$ and
\begin{equation}
{R} =  {\rm Span}\{ P[-j_1] \cdots  P[-j_k]\,1\mid
k\in {\mathbb Z}_+\cup \{0\}, \,\,\, j_1, \dots, j_k \in
{\mathbb Z}_+\}\,.
\end{equation}
1 is a highest weight vector.
This is known in physics as the {\it bosonic Fock space}.
The operators $P[j]\, (j< 0)$ ($P[j]\, (j>0)$) are the {\it creation (annihilation) operators}, while 1 is  the {\it vacuum vector}.
Define the degree operator
${\cD}: {R}\rightarrow {R}$ by \,
$
{\cD}(p_1^{m_1} p_2^{m_2}\cdot\cdot\cdot \,)\stackrel{def}{=}
(\sum_i i m_i) p_1^{m_1} p_2^{m_2}\ldots
$
The representation ${R}$ has $\cD$ eigenspace decomposition; the eigenspace with eigenvalue $N$ has a basis
$
p_1^{m_1} p_2^{m_2}\cdot\cdot\cdot
(\sum_i i m_i) = N.
$
Recall that a partition of $N$ is defined by a non-increasing sequence of non-negative integers $\nu_1\geq \nu_2 \geq \ldots$ such that $\sum_\ell \nu_\ell = N$. One can represent $\nu$ as $(1^{m_1}, 2^{m_2}, \cdot\cdot\cdot)$ (where 1 appears $m_1$-times, 2 appears $m_2$-times, and so on, in the sequence). Therefore, elements of the basis corresponds bijectively to a partition $\nu$. The generating function of eigenspace dimensions, or the {\it character}
in the terminology of representation theory, is known to have the form
\begin{equation}
{\rm Tr}_{R}\, q^{\cD}\, \stackrel{def}{=}
\sum_{N\geq 0}
q^N {\rm dim}\,\{ r\in {R}\, \mid \, {\cD}r = Nr\,\}
= \prod_{n= 1}^{\infty}(1 - q^n)^{-1}\,.
\end{equation}
Let us define now the Heisenberg algebra associated with a finite dimensional $\mathbb Q$-vector space $V$ with non-degenerate symmetric bilinear form $(\, ,\, )$. Let $W = (V\otimes t\,{\mathbb Q}[t])\oplus (V\otimes t^{-1}\,{\mathbb Q}[t^{-1}])$, then define a skew-symmetric bilinear form on $W$ by
$(r\otimes t^i,\,s\otimes t^j) = i \delta_{i+j, 0} (r, s)$.
The Heisenberg algebra associated with $V$ can be defined as follows: take the quotient of the free algebra $A(W)$ divided
by the ideal $\mathcal I$ generated by $[r,\, s] - (r,\,s)1\,\, (r, s\in W)$. It is clear that when $V = {\mathbb Q}$ one has the above Heisenberg algebra. For an orthogonal basis
$\{ r_j\}_{j=1}^p$ the Heisenberg algebra associated with $V$ is isomorphic to the tensor product of $p$-copies of the above Heisenberg algebra.

Let us consider next the {\it super}-version of the Heisenberg algebra, known as the super-Heisenberg algebra. The initial data are constituted by a vector space,
$V$, with decomposition $V = V_{\rm even}\oplus V_{\rm odd}$,
and a non-degenerate bilinear form satisfying
$(r, s) = (-1)^{|r||s|}(r, s)$. In this formula, $r, s$ are either elements of $V_{\rm even}$ or $V_{\rm odd}$, while $|r| = 0$ if
$r \in V_{\rm even}$ and $|r|=1$ if $r\in V_{\rm odd}$.
As above, we can define $W$, the bilinear form on $W$, and $A(W)/{\mathcal I}$, where now we replace the Lie bracket
$[\, ,\,]$ by the super-Lie bracket. In addition, to construct the free-super Lie algebra in the tensor algebra, we set
$
(r\otimes t^i,\,s\otimes t^j) = (r \otimes t^i)(s \otimes t^j)
+ (s \otimes t^j)(r \otimes t^i)
$
for $r, s \in V_{\rm odd}$.
By generalizing the representation on the space of polynomials of infinite many variables one can get a representation of the super-Heisenberg algebra on the symmetric algebra ${R}= S^*(V \otimes t\, {\mathbb Q} [t])$ of the positive degree part $V\otimes t\, {\mathbb Q}\,[t]$. As above, we can define the degree operator $\cD$. The following character formulas hold:
\begin{eqnarray}
{\rm Tr}_{R}\, q^{\cD} & = & \prod_{n = 1}^{\infty}
\frac{(1 + q^n)^{{\rm dim}\, V_{\rm odd}}}
{(1 - q^n)^{{\rm dim}\, V_{\rm even}}}
=
\frac{\cR(\sigma = 1-it+ i\eta(\tau))^{{\rm dim}\, V_{\rm odd}}}
{\cR(s = 1-it)^{{\rm dim}\, V_{\rm even}}}\,,
\label{trace-usual}
\\
{\rm STr}_{R}\, q^{\cD} & = & \prod_{n = 1}^{\infty}
(1 - q^n)^{{\rm dim}\, V_{\rm odd} - {\rm dim}\, V_{\rm even}}
= \cR(s = 1-it)^{{\rm dim}\, V_{\rm odd} - {\rm dim}\, V_{\rm even}}\,,
\end{eqnarray}
where we have counted the odd degree part by $-1$ and replaced the usual trace by the super-trace.\footnote{In the case when $V$ has the one-dimensional odd degree part only
(the bilinear form is $(r, r) =1$ for a nonzero vector $r\in V$) and the above condition is not satisfied, we can modify the definition of the corresponding super-Heisenberg algebra by changing the bilinear form on $W$ as $(r\otimes t^i, r\otimes t^j) = \delta_{i+j, 0}$. The resulting algebra is termed an {\it infinite dimensional Clifford algebra}. The above representation $R$ is the {\it fermionic Fock space} in physics and it can be modified as follows: the representation of the even degree part was realized as the space of polynomials of infinity many variables; the Clifford algebra is realized on the exterior algebra
${R} = \Lambda^*(\bigoplus_j{\mathbb Q} dp_j)$ of a vector space with a basis of infinitely many vectors. For $j>0$ we define $r\otimes t^{- j}$ as an exterior product of $dp_j$, and $r\otimes t^j$ as an interior product of $\partial/\partial p_j$.
}

If we consider the generating function of the Poincar\'{e} polynomials associated with sets of points we get the character of the Fock space representation of the Heisenberg algebra.
This is, in general, the integrable highest weight representation of the corresponding affine Lie algebra and is known to have modular invariance, as was proven in {\rm \cite{Kac2}}. This occurrence is naturally explained through the relation to partition functions of conformal field theory on a torus. In this connection, the affine Lie algebra has a close relationship to conformal field theory.

\subsection{One-dimensional higher variety}
\label{Variety}
Let us consider the $N$-fold symmetric product ${\mathfrak S}^NX$ of a K\"{a}hler manifold $X$, that is the ${\mathfrak S}^N X = [X^N/{\mathfrak S}_N] := \underbrace{X\times\cdots\times X}_N /{\mathfrak S}_N$ orbifold space,
${\mathfrak S}_N$ being the symmetric group of $N$ elements.
Objects of the category of the {\rm orbispace} $[X^N/{\mathfrak S}_N]$ are  the $N$-tuples $(x_1,\ldots,x_N)$ of points in $X$;  arrows are elements of the form $(x_1,\ldots,x_N; \sigma)$, where $\sigma \in {\mathfrak S}_N$. In addition, the arrow $(x_1,\ldots,x_N; \sigma)$ has as its source $(x_1,\ldots,x_N)$, and as its target $(x_{\sigma(1)},\ldots,x_{\sigma(N)})$. This category is a groupoid for the inverse of $(x_1,\ldots,x_N; \sigma)$ is $(x_{\sigma(1)},\ldots,x_{\sigma(N)}; \sigma^{-1})$.
(The orbispace as a groupoid has been describes in \cite{Kontsevich, Moerdijk}).
For a one-dimensional higher variety (i.e. for a surface) the following results hold:
\begin{itemize}
\item{}
For a Riemann surface (${\rm dim}\,X = 1$) ${\mathfrak S}^NX$ and $X^{N}$
are isomorphic under the Hilbert-Chow morphism.
\item{}
If $X$ is a nonsingular quasi-projective surface, the Hilbert-Chow morphism $\pi: \, X^{[N]}\rightarrow {\mathfrak S}^NX$ gives a resolution of the singularities of the symmetric product ${\mathfrak S}^NX$ \cite{Fogarty}. In particular, $X^{[N]}$ is a nonsingular quasi-projective variety of dimension $2N$.
\item{}
If $X$ has a symplectic form, $X^{[N]}$ has also a symplectic form.
For $N= 2$ this has been proven in \cite{Fujiki}, for $N$ general in \cite{Beauville}.
\item{} The generating function of the Poincar\'{e}
polynomials $P_r(X^{[N]})$ of $X^{[N]}$ is given by
\begin{eqnarray}
\!\!\!\!\!\!\!\!\!\!\!
\sum_{N=0}^{\infty}q^N\, P_r(X^{[N]})
& = &
\prod_{n=1}^{\infty} \frac{(1+r^{2n-1}q^n)^{b_1(X)}(1+
r^{2n+1}q^n)^{b_3(X)}}
{(1-r^{2n-2}q^n)^{b_0(X)}(1-r^{2n}q^n)^{b_2(X)}
(1-r^{2n+2}q^n)^{b_4(X)}}
\nonumber \\
& = &
\frac{\prod_{j=1, 2}\,\,\,\, \cR(\sigma = \xi_{2j-1}(1-it)+
i\eta (\tau))^{b_{2j-1}(X)}}
{\!\!\!\!\!\!\!\!\!\!\!\!\!\!\!\!\!\!\!\!\!\!
\prod_{j=1, 2, 3} \cR(s = \xi_{2j-2}(1-it))^{b_{2j-2}(X)}}
\,,
\label{GFP}
\end{eqnarray}
where $\xi_{2j-1} = j-1/2, \, \xi_{2j-2} = j-1$
and $r = \exp (\pi i\tau$).

\end{itemize}

\subsection{Equivariant K-theory, wreath products}
\label{K-theory}
We study here a direct sum of the equivariant K-groups $\cF_{\Gamma}(X):= \oplus_{N\geq 0} \underline{K}_{\Gamma_N}(X^N)$ associated with a topological $\Gamma$-space \cite{Wang}.
$\Gamma$ is a finite group and the wreath (semi-direct) product
$\Gamma_N\rtimes {\mathfrak S}_N$ acts naturally on the $N$th Cartesian product $X^N$. One can calculate the torsion free part of $K^\bullet_\Gamma(Y)$ (where $\Gamma$ acts on $Y$ and $\Gamma$ is a finite group) by localizing on the prime ideals of $R(\Gamma)$, the representation ring of $\Gamma$ (for details, see \cite{Segal})
$
\underline{K}^\bullet_\Gamma(Y)
\cong \bigoplus_{\{\gamma\}} \underline{K}^\bullet(Y^\gamma)^{\Gamma_\gamma},
$
where
$
\underline{K}_{\Gamma_N}(X^N)\equiv K_{\Gamma_N}(X^N)\otimes {\mathbb C}.
$
Here $\{\gamma\}$ runs over the conjugacy classes of elements in $\Gamma$, $Y^\gamma$ are the fixed point loci of $\gamma$ and $\Gamma_\gamma$ is the centralizer of $\gamma$ in $\Gamma$. The fixed point set $\{X^N\}^\gamma$ is isomorphic to $X^N = X^{\sum_nN_n}$, $\gamma \in {\mathfrak S}_N$, and $\Gamma_\gamma\cong \prod_n{\mathfrak S}_{N_n}\ltimes ({\mathbb Z}/n)^{N_n}$.
The cyclic groups ${\mathbb Z} / n$ act trivially in $K^\bullet(X^{N})$ and therefore the following decomposition for the $\cS_N$-equivariant $K$-theory holds \cite{Lupercio}
\begin{equation}
\underline{K}^\bullet({\mathfrak S}^NX) \cong \bigoplus_{\{\gamma\}} \underline{K}^\bullet(({\mathfrak S}^N)^\gamma))^{\Gamma_\gamma} \cong \bigoplus_{\sum nN_n= N} \bigotimes_n \underline{K}^\bullet({\mathfrak S}^{N_n})^{\mathfrak{S}_{N_n}}\,.
\end{equation}

As an example of such K-group we here analyze the group $K^{-}_{{\widetilde H}\Gamma_N}(X^N)$ which has been introduced in \cite{Wang}. The semi-direct product $\Gamma_N$ can be extended to the action of a larger finite supergroup
${\widetilde H}\Gamma_N$, which is a double cover of the semi-direct product $(\Gamma\times {\mathbb Z}_2)^N\rtimes
{\mathfrak S}_N$. The category of ${\widetilde H}\Gamma_N$-equivariant spin vector superbundles over $X^N$ is the category of $\Gamma_N$-equivariant vector bundles $E$ over $X^N$ such that $E$ carries a supermodule structure with respect to the complex Clifford algebra of rank $N$.\footnote{A fundamental example of ${\widetilde H}\Gamma_N$-vector superbundles over $X^N$ ($X$ compact) is the following:
given a $\Gamma$-vector bundle $V$ over $X$, consider the vector superbundle $V\oplus V$ over $X$ with the natural ${\mathbb Z}_2$-grading. One can endow the $N$th outer tensor product $(V\oplus V)^{\boxtimes N}$ with a natural ${\widetilde H}\Gamma_N$-equivariant vector superbundle structure over $X^N$.
}

It has been shown \cite{Wang} that the following statements hold:
{(i)} The direct sum $
\cF^{-}_{\Gamma}(X):= \oplus_{N=0}^{\infty} \underline{K}_{{\widetilde H}\Gamma_N}(X^N)
$
carries naturally a Hopf algebra structure. {(ii)}
It is isomorphic to the Fock space of a twisted Heisenberg superalgebra (in this section {\it super} means
${\mathbb Z}_2$-graded) associated with
$
K^{-}_{{\widetilde H}\Gamma_N}(X)\cong K_{\Gamma}(X).
$
{(iii)} If $X$ is a point, the K-group
$
\underline{K}_{{\widetilde H}\Gamma_N}(X^N)
$
becomes the Grothendieck group of spin supermudules of
${\widetilde H}\Gamma_N$.

Such a twisted Heisenberg algebra has played an important role in the theory of affine Kac-Moody algebras \cite{Frenkel}. The structure of the space $\cF_{\Gamma}^{-}(X)$ under consideration can be modeled on the ring $\Omega_{\mathbb C}$ of symmetric functions with a linear basis given by the so-called Schur $Q$-functions (or equivalently on the direct sum of the spin representation ring of ${\widetilde H}\Gamma_N$ for all $N$). The graded dimension of the ring $\Omega_{\mathbb C}$ is given by the denominator $\prod_{n=0}(1-q^{2n-1})^{-1}$. On the basis of G\"{o}ttsche's formula \cite{Gottsche} it has been conjectured \cite{Vafa1} that the direct sum $\cH(S)$ of homology groups for Hilbert scheme $S^{[N]}$ of $N$ points on a (quasi-)projective surface $S$ should carry the structure of the Fock space of a Heisenberg algebra, which was realized subsequently in a geometric way \cite{Nakajima,Grojnowski}. Parallel algebraic structures such as Hopf algebra, vertex operators, and Heisenberg algebra as part of vertex algebra structures \cite{Borcherds1,Frenkel} have naturally showed up in $\cH(S)$ as well as in $\cF_{\Gamma}(X)$. If $S$ is a suitable resolution of singularities of an orbifold $X/\Gamma$, there appears close connections between $\cH(S)$ and $\cF_{\Gamma}(X)$ \cite{Wang}. In fact the special case of $\Gamma$ trivial is closely related to the analysis considered in \cite{Dijkgraaf}. It would be interesting to find some applications of results discussed above in string theory.

\noindent
{\bf The generating function.}
The orbifold Euler number ${\bf e}(X,\G)$ was introduced in
\cite{Dixon} in the study of orbifold string theory and it
has been interpreted as the
Euler number of the equivariant K-group $K_{\Gamma}(X)$
\cite{Atiyah}. Define the Euler number of the generalized
symmetric product to be the difference
$
{\bf e}(X^N,{\widetilde H}\Gamma_N) := \dim K^{-,0}_{{\widetilde H}\Gamma_n}(X^N) - \dim
K^{-,1}_{{\widetilde H}\Gamma_N}(X^N)\,,
$
the series $\sum_{N=0}^{\infty} q^N {\bf e}(X^N,{\widetilde H}\Gamma_N)$ can be written in terms of spectral functions:
\begin{eqnarray}
\sum_{N=0}^{\infty} q^N {\bf e}(X^N,{\widetilde H}\Gamma_N)
& = &
\prod_{ n=1}^{\infty} (1-q^{2n-1})^{ -{\bf e}(X,\G)}
=
\left[q^{-\frac{25}{24}}(q-1)f_3(q)
\right]^{{\bf e} (X, \Gamma)}
\nonumber \\
& = &
\cR(s= 1/2 - (1/2)it)^{-{\bf e}(X, \Gamma)}
\,.
\end{eqnarray}

One can give an explicit description of ${\cF}^{-}_{\Gamma}(X)$
as a graded algebra.
Indeed, the following statement holds \cite{Wang}:             %
As a $(\Z_+ \times \Z_2)$-graded algebra, ${\cF}^-_{\G}(X, q)$
is isomorphic to the supersymmetric algebra
$ {\mathfrak S} \left( \bigoplus_{ N=1}^{\infty} q^{2N-1} \underline{K}_\G(X)\right)$. In particular,
\begin{eqnarray}
\dim_q {\cF}^{-}_{\Gamma}(X) & = &
\prod_{ n=1}^{\infty}
\frac{(1 + q^{2n-1})^{ \dim K^1_\Gamma (X)} }
{(1 - q^{2n-1})^{ \dim K^0_\Gamma (X)}}
\nonumber \\
& = &
\frac{\left[\cR(\sigma=1/2 - (1/2)it + (1/2)i\eta(\tau))
\right]^{{\rm dim} K^1_\Gamma (X)}}
{\left[\cR(s=1/2 -(1/2)it)\right]^{{\rm dim} K^0_\G (X)}},
\end{eqnarray}
where the supersymmetric algebra is equal to the tensor product of
the symmetric algebra
$ {\mathfrak S} \left( \bigoplus_{ N=1}^{\infty} q^{2N-1} \underline{K}^0_\Gamma (X)\right)$ and the exterior algebra
$\Lambda \left( \bigoplus_{ N=1}^{\infty} q^{2N-1}
\underline{K}^1_\G(X) \right)$.
In the case when $X_{pt}$ is a point we have
\begin{equation}
\sum_{N \geq 0} q^N \dim \cF^{-}_{\Gamma}(X_{pt})
= \prod_{ n=1}^{\infty} (1 -q^{2n-1})^{ -|\Gamma_*|} =
\left[\cR(s=1/2 -(1/2)it)\right]^{ -|\Gamma_*|}\,.
\label{gamma00}
\end{equation}
In (\ref{gamma00}) $\Gamma$ is a finite group with $r+1$ conjugacy classes; $\Gamma^* := \{\gamma_j\}_{j=0}^r$ is the set of complex irreducible characters, where $\gamma_0$ denotes the trivial character. By $\Gamma_*$ we denote the set of conjugacy classes.

\section{Conclusions}
\label{Conclusions}
Having advocated in this paper the basic role of modular forms and spectral functions with their connection to Lie algebra cohomologies anf K-theory methods, we are now, in concluding, naturally led to other crucial problems, related to the deformation procedure. For instance, one might ask whether the form of the partition functions could be algebraically interpreted by means of infinitesimal deformations of the corresponding Lie algebras. No doubt this analysis will require a new degree of mathematical sophistication. Perhaps all the concepts of what could eventually be called the ``deformation theory of everything" might be possibly tested in the case of associative algebras, which are algebras over operads \cite{Kontsevich00}. In many examples dealing with algebras over operads, arguments of the universality of associative algebras are called forth. This strongly suggests that a connection between the deformation theory (deformed partition functions) and algebras over operads might exist. We expect to be able to discuss this key problem in forthcoming work.

\begin{acknowledgement}
AAB would like to acknowledge the Conselho Nacional
de Desenvolvimento Cient\'ifico e Tecnol\'ogico (CNPq, Brazil) and Funda\c cao Araucaria (Parana, Brazil) for financial support.
EE's research has been partly supported by MICINN (Spain), contract PR2011-0128 and projects FIS2006-02842 and FIS2010-15640, by the CPAN Consolider Ingenio Project, and by AGAUR (Generalitat de Ca\-ta\-lu\-nya), contract 2009SGR-994.
\end{acknowledgement}

\end{document}